\documentstyle[12pt]{article}
\title{\bf Four Quantum Conservation Laws on Black Hole Equilibrium 
                 Radiation Process and Quantum Black Hole Entropy } 
\author{S. Q. Wu\thanks{E-mail: emu@iopp.ccnu.edu.cn } and X. Cai\thanks{
                       E-mail:xcai@wuhan.cngb.com }  \\
   \footnotesize \sl Institute of Particle Physics, Hua-Zhong 
                    Normal University, Wuhan, 430079, China } 

\date{\today}
\begin{document}
\maketitle
\begin{quote}

The classical first law of thermodynamic for Kerr-Newmann black hole ( KNBH ) 
is generalized to that in quantum form on event horizon. Then four quantum 
conservation laws on the KNBH equilibrium radiation process are derived. 
As a by-product, Bekenstein-Hawking's relation $ {\cal{S}}= {\cal{A}}/4 $ 
is exactly established. It can be argued that the classical entropy of black 
hole arise from the quantum entropy of field quanta or quasi-particles inside 
the hole. 

\vskip 0.2cm
  PACS number(s): 04.60.+n, 04.20.Me, 97.60.L

\end{quote}
\vskip 0.4cm

It has been nearly one quarter century since Benkenstein$^{1}$ and 
Hawking$^{2}$ first showed that the entropy of a black hole is one quarter 
of its surface area. Despite considerable effort$^{3}$ about the quantum$^{4}$, 
dynamic$^{5}$, or statistical$^{6}$ origin of black hole thermodynamics, 
however, the exact source and mechanism of the Benkenstein-Hawking black 
hole entropy remain unclear$^{7}$.

By using the brick wall model, 't Hooft$^{8}$ identified the black hole
entropy with the entropy of a thermal gas of quantum field excitations
outside the event horizon. Frolov and Novikov$^{4}$ argued that the black 
hole entropy can be obtained by identifying the dynamical degrees of freedom 
of a black hole with the states of all fields which are located inside the 
black hole. Such two opinions differ from each other. A black hole acts as
a classical object, but its true microscopic structure still remains unknown. 

In this paper, we first derive the thermal spectrum and entropy of a massive 
complex scalar field on Kerr-Newmann black hole ( KNBH ) background. From this 
quantum entropy, we propose a quantum first law of black hole thermodynamics. 
Then we consider a system in which a KNBH is in equilibrium with this scalar 
field. We could regard this system as a two-phase equilibrium one which is
consisted of radiant phase and black hole phase, with its interface being 
the event horizon. Further the black hole phase can be thought as a condensed 
state which is made up of field quanta inside the hole, while the radiant 
phase being a excited state of the same field quanta outside the hole. In this
thermodynamical picture, the event horizon acts as a membrane$^{9}$. Using 
thermodynamical method, we obtain four conservation laws on black hole 
equilibrium radiation process for energy, charge, angular momentum and entropy 
respectively. The total quantities of these quantum numbers of the whole system 
are conserved in this stationary thermal equilibrium radiation process. By 
identifying this complex scalar field with quasi-particles excited by the hole,
we propose that the classical entropy of a black hole originate microscopically 
from the entropy of quanta which constitute the hole.

The general stationary axial solution to Einstein equation is a rotating 
charged black hole ( KNBH ) described by three parameters : mass $ M $, charge 
$ Q $, specific angular momentum $ a=J/M $. So we deal with a source-less 
charged massive scalar field with mass $ \mu $ and charge $ q $ on this 
background in the non-extreme case ( $ 0 < \varepsilon = \sqrt{M^2-a^2-Q^2} 
\leq M ) $. ( In Planck units system $ G=\hbar=c=k_B=1 $ ).

In the Boyer-Lindquist coordinates, a complex scalar wave function $ \Psi $ 
has a solution of variables separable form$^{10}$: 
\begin{equation}
 \Psi(t,r,\theta,\varphi)=R(r)S(\theta)e^{(im\varphi-\omega t)} 
\end{equation}

\noindent
here the angular wave function $ S(\theta) $ is an ordinary spheroidal with 
spin-weight $ s=0 $ which satisfies Legendre wave equation$^{11}$:
\begin{equation}
 \frac{1}{\sin\theta}\partial_{\theta}[\sin\theta\partial_{\theta}S(\theta)]
 +[\lambda-\frac{m^2}{\sin^2\theta}-(ka)^2\sin^2\theta]S(\theta)=0, 
\end{equation}

\noindent
while the radial wave function $ R(r) $ is a modified generalized spheroidal 
wave function with an imaginary spin-weight, which satisfies the following 
"modified" generalized spin-weight spheroidal wave equation of imaginary 
number order$^{12,13}$: 
\begin{eqnarray}\nonumber
 \partial_r [(r-r_+)(r-r_-)\partial_r R(r)]+[k^2(r-r_+)(r-r_-)
 +2(A\omega-M{\mu}^2)(r-M) \\
 +\frac{[A(r-M)+\varepsilon B]^2}{(r-r_+)(r-r_-)}
 -(2{\omega}^2-{\mu}^2)(2M^2-Q^2)-2qQ M-\lambda]R(r)=0,
\end{eqnarray}

\noindent
where $ \lambda $ is a separation constant, $ r_{\pm}= M \pm \varepsilon, k^2
={\omega}^2-{\mu}^2, A=2M\omega-qQ, \varepsilon B=\omega(2M^2-Q^2)-qQM-ma $. 

When considering thermal radiation of the KNBH, we need the asymptotic
solutions of the radial function $ R(r) $ at its event horizon $ r=r_+ $. 
In fact, the radial equation has two solutions whose indices at its regular
singularities $ r=r_{\pm} $ are $ \pm iW $, where $ W $ is given blow. The two 
asymptotic solutions are:
\begin{equation}
 R(r) \sim (r-r_+)^{\pm i W}, \hskip 0.5cm 
 {\rm when} \hskip 0.5cm r \rightarrow r_+
\end{equation}

According to the analytical continued method suggested by Damour-Ruffini$^{14}$, 
these two solutions differ by a extra factor $ e^{2\pi W} $. Then, it is easy 
to obtain a thermal radiation spectrum$^{15}$:
\begin{equation}
  <N>=\frac{1}{e^{4\pi W}-1}, \hskip 0.5cm 
  W=\frac{\omega-m\Omega-q\Phi}{2\kappa} .
\end{equation}

\noindent
where surface gravity $ \kappa=(r_+-M)/{\cal{A}} $, angular velocity 
$ \Omega=a/{\cal{A}} $, electrical potential $ \Phi=Qr_+/{\cal{A}} $, 
reduced horizon area $ {\cal{A}}=r_+^2+a^2 $. 

In fact, quantum number $ W $  is the entropy of quantized scalar fields on
KNBH background whose energy satisfies relation:
\begin{equation}  
  \omega=2\kappa  W+m\Omega+q\Phi=2\kappa W+\omega_+. 
\end{equation}

This relation demonstrates that a KNBH can have two radiation mechanisms:
superradiant mode ( $ \omega < \omega_+ $ ) and Hawking mode ( $ \omega
> \omega_+ ) ^{16}$. In thermal equilibrium radiation process, the hole's 
surface gravity, angular velocity and electrical potential remain unchanged. 
By differentiating the energy relation of Eq.( 6 ), we have the following 
first law of quantum thermodynamics in differential form:
\begin{equation}
 \Delta\omega=2\kappa\Delta W+\Omega\Delta m+\Phi\Delta q.
\end{equation}

Let us consider a KNBH in thermal equilibrium with a complex scalar field 
at temperature $ {\cal{T}}=\kappa/2 $. This system can be thought as a 
two-phase equilibrium system which is consisted of black hole phase and 
radiant phase. The black hole phase is a liquid state condensed from ground 
state field quanta inside the hole, while the radiant phase being a gas phase
constituted of excited state field quanta outside the hole. These two phases,  
as is being interfaced by the event horizon, are two different kinds of states 
of the same field quanta. In this thermodynamical system, there exists detailed 
balance process$^{17}$, that is, the number of quanta emitted by the hole is 
equal to that absorbed by it. From conditions of thermodynamical equilibrium 
on event horizon :
$$ \kappa_{r>r_+}=\kappa_{r<r_+}, \Omega_{r>r_+}=\Omega_{r<r_+},
 \Phi_{r>r_+}=\Phi _{r<r_+}, $$

\noindent
combining Eq.( 7 ) with the following classical first law of black hole 
thermodynamics in differential form$^{18}$: 
\begin{equation}
 \Delta M=\frac{\kappa}{2}\Delta{\cal{A}}+\Omega\Delta J+\Phi\Delta Q,
\end{equation}

\noindent
we can deduce four quantum conservation laws for energy, angular momentum, 
charge and entropy respectively.
\begin{eqnarray} 
 \hskip -2cm {\rm Energy}: \hskip 2.9cm \Delta M &=& \Delta\omega,  \\
 \hskip -2cm {\rm Angular \hskip 3pt Momentum}: \hskip 1cm 
  \Delta J &=& \Delta m, \\
 \hskip -2cm {\rm Charge}: \hskip 3cm \Delta Q &=& \Delta q, \\
 \hskip -2cm {\rm Entropy}: \hskip 2.5cm \frac{1}{4}\Delta{\cal{A}} &=&
  \Delta W. 
\end{eqnarray}

When considering all modes of field quanta, the above conservation relations ( 
9-12 ) must include sum respect to all possible modes of field configurations.
Eqs.( 9-12 ) indicate that a KNBH has discrete increment of energy, angular 
momentum, charge and entropy. That is, when a black hole emits particles, its
energy, charge, angular momentum and entropy are carried away by these quanta
, and vice versa. In a stationary thermal equilibrium radiation process, it is 
reasonable physically conceived that what the hole gains meets with the ends
of that the radiation loses. Thus, the total quantities of energy, charge, 
angular momentum and entropy remain unchanged in this thermodynamic process.

Further, combining Eq.( 6 ) with integral Smarr formulae$^{18}$:
\begin{equation}
 M=\kappa{\cal{A}}+2J\Omega+Q\Phi,
\end{equation}

\noindent
we can obtain a special quantum state $ m=J, \omega=M/2, q=Q/2, W={\cal{A}}/4 $. 
This demonstrates that a quantum KNBH is a collection of all possible quasi
-particles inside the hole. As quantum number $ m, \omega, q, W $ are discrete 
numbers, not only the parameters $ J, M, Q, {\cal{A}} $ but also $ \Delta J, 
\Delta M, \Delta Q, \Delta{\cal{A}} $ take discrete values. So, a KNBH could 
be thought as a condensed phase consisted of all possible modes of bosonic 
field quanta.  

In fact, Eq.( 12 ) is a generalized second thermodynamic law in quantum form.
By integrating this equation, we obtain quantum black hole entropy:
\begin{equation}
   W=\frac{1}{4}{\cal{A}}+C.
\end{equation} 

As Bekenstein-Hawking's classical black hole entropy$^{1,2 } : \rm{S}=\rm{A}/4
=\pi{\cal{A}} $, the quantum entropy $ W $ is equal to the reduced entropy 
$ W={\cal{S}}=\rm{S}/(4\pi) $, so we have Bekenstein-Hawking relations ( Choose
constant ( $ C=0 $ ):
\begin{equation}      
  {\cal{S}}=W=\frac{1}{4}{\cal{A}}.
\end{equation}

Eq.( 15 ) shows that Bekenstein-Hawking black hole entropy is equal to 
quantum entropy of a complex scalar field. In other words, the classical 
entropy of black holes originates statistically from quantum entropy of 
quantized fields.

The point of our view that a black hole is being made up of radiation field
quanta can be simply illustrated by a soap bubble thermodynamical model. Let 
us consider a sphere with radius $ r=2M $, mass $ M $ in vacuum, this sphere 
contains uniform radiation inside it. A condition that this bubble doesn't 
break down is that its pressure $ p $ must be equal to its surface tension 
$ \sigma $ being divided by its mass $ M $, that is, $ p=\sigma /M=\rho /3 $, 
where $ \rho=3/(32\pi M^2) $ being the mean density of radiations in this
sphere. So we have a relation $ \sigma=1/(32\pi M) $. Then the surface gravity $ \kappa=8\pi \sigma=1/(4M) $, which 
is exactly equal to the surface gravity in a Schwarzschild black hole case. 
Thus, the radiant pressure can contend with the gravity of a sphere symmetric
black hole. ( Note: The factor $ 8\pi $ rises from Einstein coupled coefficient, 
and $ \rho=3p $ is state equation of photon radiations in our argument ). On 
the other hand, the energy of radiations inside the sphere can't take arbitrary 
values, it must be multiple of ground state energy $ 8\pi M $ which is due to 
Heisenberg uncertain principle.

In summary, we regard a Kerr-Newmann black hole as a classical two-phase 
thermodynamical system. Using two-phase thermodynamical equilibrium condition, 
we derive four quantum conservation laws on black hole equilibrium radiation 
process. The total energy, total charge, total angular momentum, total entropy 
of the whole system are conserved in this process. By identifying a KNBH with 
a collection of condensed ground state quanta, we propose that the classical 
entropy of a black hole originate microscopically from quantum entropy of 
ground state quanta inside the hole.  
 
\vskip 4pt

This work is supported partly by the NNSF and Hubei-NSF in China.

\begin{enumerate}
 \item J. D. Bekenstein, Phys. Rev. D 7, 2333 (1973); D 9, 3292 (1974).
 \item S. W. Hawking, Commun. Math. Phys. 43, 199 (1975).
 \item J. Benkenstein, {\sl Do we understand black hole entropy ?}, 
  gr-qc/9409015.
 \item L. Bombelli, R. Koul, J. Lee and R. Sorkin, Phys. Rev. D34, 373
  (1986).
 \item V. Frolov and I. Novikov, Phys. Rev. D 48, 4545 (1993);
  V. Frolov, Phys. Rev. Lett. 74, 3319 (1995).
 \item W. H. Zurek and K. S. Thorne, Phys. Rev. Lett. 50, 2171 (1985).
 \item M. H. Lee and J. Kim , hep-th/9604130; hep-th/9602129;
  J. Ho, W. T. Kim, Y. J. Park, gr-qc/9704032. 
 \item G. 't Hooft, Nucl. Phys. B 256, 727 (1985);
  L. Susskind and J. Uglum, Phys. Rev. D 50, 2700 (1994). 
 \item K. S. Thorne, {\sl et al., Black Holes: The Membrane Paradigm},
  edited by K. S. Thorne, R. H. Price and D. A. Macdonald, Yale University
  Press, 1986; 
  M. Maggiore, Nucl. Phys. B 429, 205 (1994). 
 \item B. Carter, Phys. Rev. 174, 1559 (1968).
 \item P. Moon and D. E. Spencer, {\sl Field Theory Handbook}, Springer
  -Verlag, New York, 1971, 2nd version;
  P. M. Morse, H. Feshbach, {\sl Methods of Theoretical Physics}, McGraw-Hill, 
  New York, 1953;
  {\sl Handbook of Mathematical Functions}, edited by M. Abramowitz \& I. A. 
  Stegun, Dover, New York, 9th version, 1972.
 \item S. Q. Wu, X. Cai, {\sl Exact Solutions to Source-less Charged Massive
 Scalar Field Equation on Kerr-Newmann Background}, submitted to J. Math. Phys.
 \item E. W. Leaver, J. Math. Phys. 27, 1238 (1986);
 \item T. Damour, R. Ruffini, Phys. Rev. D 14, 332 (1976).
 \item J. G. Bellido, hep-th/9302127;
  G. W. Gibbons and M. J. Perry, Proc. R. Soc. Lond. A. 358, 467 (1978).
 \item R. M. Ward, {\sl General Relativity}, Chicago University Press, 
  1984.
 \item D. Bekenstein and A. Meisels, Phys. Rev. D 15, 2775 (1976). 
 \item L. Smarr, Phys. Rev. Lett. 30, 71 (1973); 30, 521(E) (1973).
\end{enumerate}

\end{document}